\documentstyle[hip-artc,epsf]{article}
\recrevdate{February 24, 1999}

\def\rightmark{Relation to centrality}
\def\leftmark{R. Vogt}

\title{Relation of hard and total cross sections to 
centrality}

\authors{
{\twerm R. Vogt}\\[2.812mm]
{\normalsize 
Nuclear Science Division, Lawrence Berkeley National Laboratory\\
University of California, Berkeley, California 94720 \\
and \\
Physics Department, \\
University of California, Davis, California, 95616
}
}

\abstract{
We compare the fractions of the hard and geometric cross sections as a function
of impact parameter.  For a given definition of central collisions, we
calculate the corresponding impact parameter and the fraction of the hard
cross section contained within this cut.  We use charm quark production as a 
definite example. 
}

\begin{document}
\maketitle

In this note, revised from Ref.~\cite{evw},
standard nuclear density distributions are described and the
resulting geometrical overlap in nuclear collisions is calculated.  We then
compare the hard process cross section to the total geometric
cross section as a function of impact parameter and discuss how they are
related to the collision centrality.

A three--parameter Woods--Saxon shape is used to describe the nuclear density
distribution, \begin{eqnarray} \rho_{\rm A}(r)=\rho_0 \frac{1 + 
\omega(r/R_{\rm A})^2}{1 +
\exp((r-R_{\rm A})/z)} \, \, , \label{density} 
\end{eqnarray} where $R_A$ is the nuclear radius,
$z$ is the surface thickness, and $\omega$ allows for central irregularities. 
The electron scattering data of Ref.\ \cite{Vvv} is used where available
for $R_{\rm A}$, $z$, and $\omega$.  When data is unavailable, the
parameters $\omega =0$, $z=0.54$ fm and $R_{\rm A} = 1.19A^{1/3} -
1.61A^{-1/3}$ fm are used.  The central density $\rho_0$ is found from 
the normalization $\int d^3r \rho_{\rm A}(r) = A$.  
For results with other nuclear
shape parameterizations, see the appendix of Ref.\ \cite{EKL}.
The nuclear shape parameters are given in Table~\ref{shapes}.  
See also Ref.~\cite{shape}
for more detailed discussion of nuclear density distributions.

\begin{table}[htbp]
\begin{center}
\begin{tabular}{|c|c|c|c|c|} \hline
$A$  &  $R_{\rm A}$ (fm)  &  $z$ (fm)  &  $\omega$  &  $\rho_0$ (fm$^{-3}$) 
\\ \hline
16   &   2.608     &  0.513    &  -0.051    &  0.1654  \\
27   &   3.07      &  0.519    &     0.     &  0.1739  \\
40   &   3.766     &  0.586    &  -0.161    &  0.1699  \\
63   &   4.214     &  0.586    &     0.     &  0.1701  \\
110  &   5.33      &  0.535    &     0.     &  0.1577  \\
197  &   6.38      &  0.535    &     0.     &  0.1693  \\ 
208  &   6.624     &  0.549    &     0.     &  0.1600  \\ \hline
\end{tabular}
\end{center}
\caption[]{ Nuclear shape parameters taken from Ref.~\cite{Vvv}.}
\label{shapes}
\end{table}

In minimum bias (impact parameter averaged) $AB$ collisions we expect the
production cross section for hard processes to grow approximately as 
\begin{eqnarray} \sigma_{\rm AB}^{\rm hard} = \sigma_{pp}^{\rm hard} 
(AB)^\alpha \, \, , \label{hard}
\end{eqnarray}  where $\alpha \equiv 1$ when no nuclear effects are included.
However, central collisions
are of the greatest interest since it is there that high energy density effects
are most likely to appear.  Central
collisions contribute larger than average values of $E_T$ to the system, in the
`tail' of the $E_T$ distribution, $d\sigma/dE_T$.  We would like to determine
which impact parameters are important in the high $E_T$ tail, {\it i.e.}\ 
what range of $b$ may be considered central.  We now define the central
fraction of the hard cross section, Eq.~(\ref{hard}), and the central fraction
of the geometric cross section.  We then discuss how the two are related.

\begin{figure}[htb]
\setlength{\epsfxsize=\textwidth}
\setlength{\epsfysize=0.4\textheight}
\centerline{\epsffile{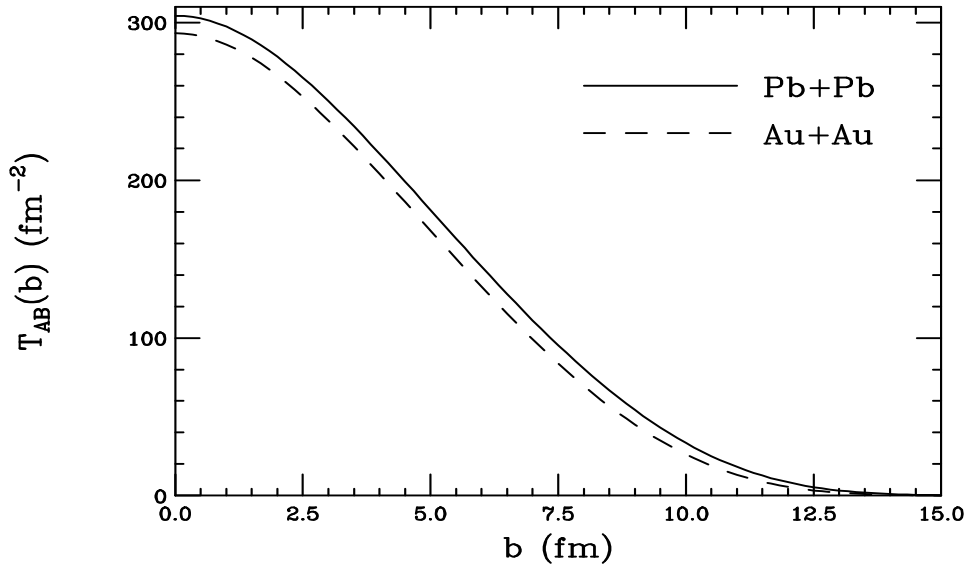}}
\caption[]{The nuclear overlap function $T_{\rm AB}(b)$ as a function of impact
parameter $b$ for Pb+Pb and Au+Au collisions. } 
\label{tab}
\end{figure}

Considering
only geometry with no nuclear effects, $\alpha = 1$ in Eq.~(\ref{hard}), 
the inclusive production cross section of hard probes increases
as 
\begin{eqnarray}
d\sigma_{\rm AB}^{\rm hard} = \sigma_{pp}^{\rm hard} T_{\rm AB}(b) d^2b
\label{abhard} \end{eqnarray} 
and the average number of hard probes  produced at impact parameter $b$ 
is $\overline N_{\rm AB}(b)
=  \sigma_{pp}^{\rm hard} T_{\rm AB}(b)$ where $T_{\rm AB}(b)$
is the nuclear overlap integral, \begin{eqnarray} T_{\rm AB}(\vec b) = \int 
d^2s \, T_{\rm A}(\vec s) \, T_{\rm B}(\vec b - \vec s) \, \, , \label{over}
\end{eqnarray} and $T_{\rm A} = 
\int dz \rho_{\rm A}(z, \vec s)$ is
the nuclear profile function.  The nuclear overlap functions for Pb+Pb and 
Au+Au collisions are shown in Fig.~\ref{tab} 
as a function of impact parameter.  
Integrating $T_{\rm AB}$ over all impact
parameters we find \begin{eqnarray} \int d^2b \, T_{\rm AB}(b) = AB \, \,
. \label{total} 
\end{eqnarray}  

\begin{figure}[htb]
\setlength{\epsfxsize=\textwidth}
\setlength{\epsfysize=0.4\textheight}
\centerline{\epsffile{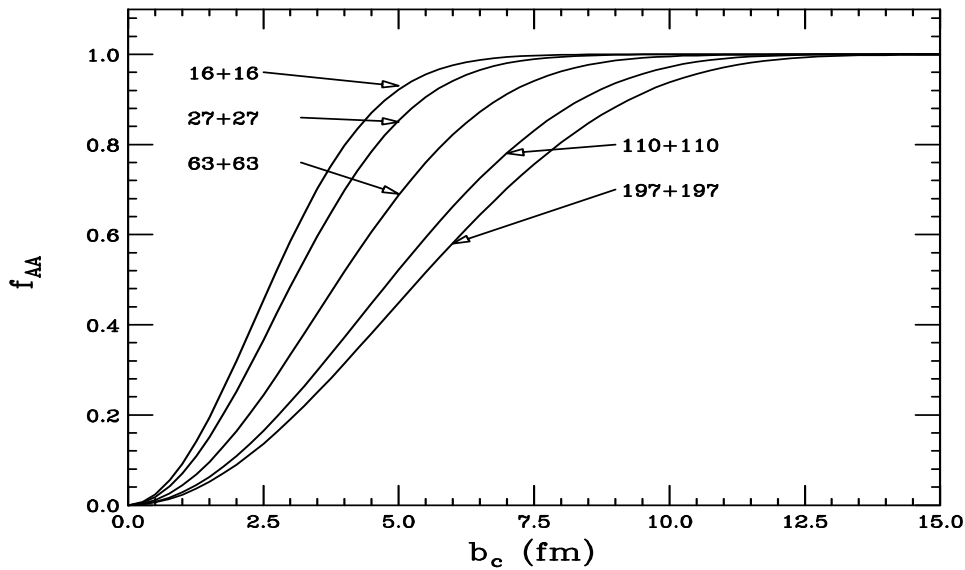}}
\caption[]{The central fraction of the hard cross section as a function of
impact parameter cut $b_c$ for several symmetric systems. } 
\label{fab}
\end{figure}

The central fraction $f_{\rm AB}$, equivalent to the fraction of the total
hard cross section,
is defined as \begin{eqnarray} f_{\rm AB} = \frac{2 \pi}{AB}
\int_0^{b_c} b \, db \, T_{\rm AB}(b) \,\, , \label{abfrac}
\end{eqnarray} where $b_c$ is 
the central impact parameter and $b<b_c$ are central.  
To make a similar `centrality cut' in $pA$
collisions, the fraction \begin{eqnarray} f_{\rm A} = \frac{2 \pi}{A} 
\int_0^{b_c} b \, db \, T_{\rm A}(b) \, \, \label{afrac}
\end{eqnarray} would be used.  
Fig.~\ref{fab}, taken from Ref.\ \cite{RV}, shows the increase of $f_{\rm AB}$ 
with $b_c$ for several symmetric, $AA$, systems.  Note that $f_{\rm AA}
\approx 1$ when $b_c \approx 2R_A$.  For example, if we choose $\sigma_{\rm
central} = 0.1 \sigma_{\rm AB}^{\rm hard}$, 
this corresponds to $b_c = 2.05$ fm in
Au+Au collisions and $b_c = 1.05$ fm in O+O collisions.  If we instead chose
$\sigma_{\rm central} = 0.01 \sigma_{\rm AB}^{\rm hard}$ 
then $b_c = 0.52$ fm in Au+Au
and $b_c = 0.33$ fm in O+O collisions.  

Note however that $f_{\rm AB}$ 
is not the fraction of the geometric cross section
which includes both hard and soft contributions.  The
geometric cross section in central collisions is found by integrating the
interaction probability over impact parameter up to $b_c$,
\begin{eqnarray} \sigma_{\rm geo}(b_c) = 2 \pi \int_0^{b_c} b\, db \, 
[1 - \exp(-T_{\rm AB} \sigma_{NN})] \, \, . \label{etgeo} \end{eqnarray}  The
nucleon-nucleon inelastic cross section, $\sigma_{NN}$, is $\approx 32$ mb at
SPS energies and grows with energy. It is expected to be $\sim 60$ mb at LHC
energies.  The fraction of the geometric cross section is \begin{eqnarray}
f_{\rm geo} = \frac{\sigma_{\rm geo}(b_c)}{\sigma_{\rm geo}} \, \, .
\label{geofrac} \end{eqnarray}
In central collisions, where $T_{\rm AB}$ is large, the impact
parameter dependence is simple, $\sigma_{\rm geo}(b_c) \propto b_c^2$.  
However, in peripheral collisions where the nuclear overlap becomes small, 
$\sigma_{\rm geo}(b_c)$
deviates from the trivial $b_c^2$ scaling.  Deviations from this scaling do
not occur until $b_c \approx 2R_A$ in symmetric systems.

\begin{figure}[htb]
\setlength{\epsfxsize=\textwidth}
\setlength{\epsfysize=0.4\textheight}
\centerline{\epsffile{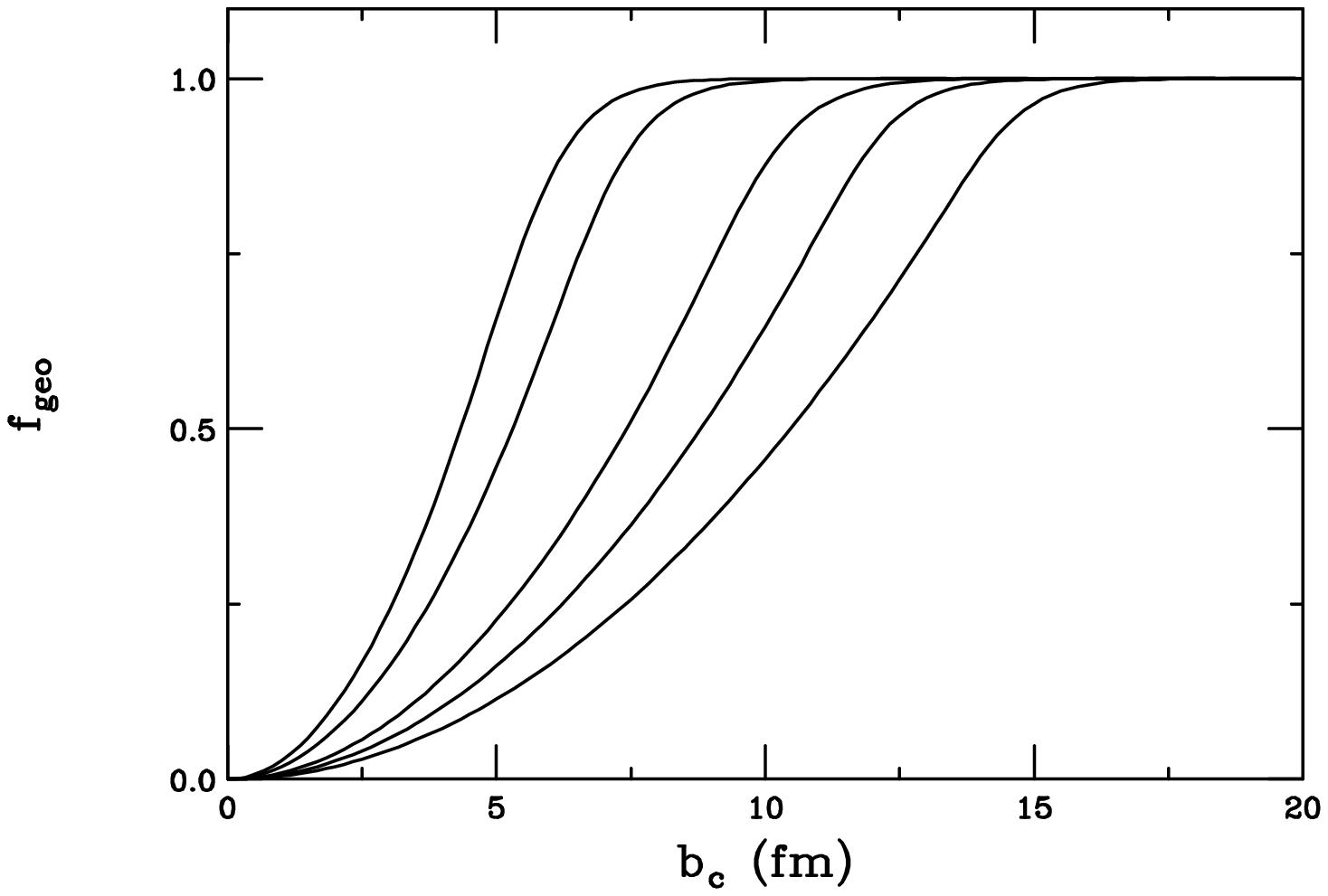}}
\caption[]{The fraction of the total geometrical cross section as a function 
of impact parameter cut $b_c$ for several symmetric systems.  From left to
right the curves are 16+16, 27+27, 63+63, 110+110, and 197+197.} 
\label{fgeo}
\end{figure}

Figure~\ref{fgeo} 
shows the numerical result, Eq.~(\ref{etgeo}), relative to the 
integral where $b_c \rightarrow \infty$, for the same systems as in
Fig.~\ref{fab}. 
We have used $\sigma_{NN} = 32$ mb in Eq.~(\ref{etgeo}).  A negligible
difference in the most peripheral collisions can be expected if 60 mb were used
instead.
The growth of the fraction of the geometric cross section is slower than that
of the hard fraction, $f_{\rm AB}$.  Indeed at $b_c \approx 2R_A$, $f_{\rm 
geo} \approx 0.75$.

The total geometrical cross section for a variety of colliding nuclei is given
in Table~\ref{geotab}.  
We have also calculated the impact parameter $b_c$ corresponding
to $f_{\rm geo} = 0.05$, 0.1, and 0.2 or the central 5\%, 10\% and 20\% of all
collisions.  The impact parameter corresponding to $f_{\rm geo} = 0.2$ is 
$b_c \approx 1.04 R_A$ when symmetric systems are considered.  In asymmetric
collisions, $b_c < R_A$ when $f_{\rm geo} = 0.2$.  If smaller centrality cuts
are imposed, the impact parameters are reduced by factors of $\sqrt{2}$ and 2
for $f_{\rm geo} = 0.1$ and 0.05 respectively.  Note that when $B \gg A$, in
the smaller nucleus is embedded in the larger with a 10\% or smaller centrality
cut. 

\begin{table}[htpb]
\begin{center}
\begin{tabular}{|c|c|c|c|c|} \hline
 & & \multicolumn{3}{c|}{$b_c$ (fm)} \\
$A+B$   &  $\sigma_{\rm geo}$ (b) & $f_{\rm geo} = 0.05$ & $f_{\rm geo} = 
0.1$ & $f_{\rm geo} = 0.2$ \\ \hline
16+16   & 1.18 & 1.37 & 1.94 & 2.74 \\
16+27   & 1.45 & 1.52 & 2.15 & 3.04 \\
16+40   & 1.75 & 1.67 & 2.36 & 3.34 \\
16+63   & 2.15 & 1.85 & 2.62 & 3.70 \\
16+110  & 2.69 & 2.07 & 2.93 & 4.14 \\
16+197  & 3.42 & 2.33 & 3.30 & 4.66 \\
16+208  & 3.59 & 2.39 & 3.38 & 4.78 \\
27+27   & 1.76 & 1.67 & 2.37 & 3.35 \\
27+40   & 2.09 & 1.82 & 2.58 & 3.65 \\
27+63   & 2.53 & 2.01 & 2.84 & 4.01 \\
27+110  & 3.11 & 2.22 & 3.15 & 4.45 \\
27+197  & 3.89 & 2.49 & 3.52 & 4.98 \\
27+208  & 4.08 & 2.55 & 3.60 & 5.09 \\
40+40   & 2.45 & 1.98 & 2.79 & 3.95 \\
40+63   & 2.93 & 2.16 & 3.05 & 4.32 \\
40+110  & 3.55 & 2.38 & 3.36 & 4.76 \\
40+197  & 4.38 & 2.64 & 3.73 & 5.28 \\
40+208  & 4.58 & 2.70 & 3.82 & 5.40 \\
63+63   & 3.46 & 2.34 & 3.32 & 4.69 \\
63+110  & 4.14 & 2.56 & 3.63 & 5.13 \\
63+197  & 5.04 & 2.83 & 4.00 & 5.66 \\
63+208  & 5.25 & 2.89 & 4.09 & 5.78 \\
110+110 & 4.86 & 2.78 & 3.93 & 5.56 \\
110+197 & 5.82 & 3.04 & 4.30 & 6.09 \\
110+208 & 6.06 & 3.10 & 4.39 & 6.21 \\
197+197 & 6.88 & 3.31 & 4.68 & 6.62 \\ 
197+208 & 7.13 & 3.37 & 4.76 & 6.74 \\
208+208 & 7.39 & 3.43 & 4.85 & 6.86 \\  \hline
\end{tabular}
\end{center}
\caption[]{ Values of the geometric cross section and the impact parameter at
which $f_{\rm geo} = 0.05$, 0.1 and 0.2 respectively for 
several colliding systems. }
\label{geotab}
\end{table}

For the same systems, Table~\ref{tabtab} 
gives the value of the nuclear
overlap at $b=0$, $T_{\rm AB}(0)$, and the fraction of the hard cross section,
$f_{\rm AB}$, corresponding to 5\%, 10\%, and 20\% of the geometric cross
section, calculated with the $b_c$ values given in Table~\ref{geotab}.
For example, the central 10\% of the total
geometric cross section is obtained when $b_c \sim 4.7$ 
fm in Au+Au collisions, more than twice the
corresponding impact parameter for the same percentage of the hard cross
section.  In this case, 10\% of the geometric cross section encompasses
$\approx 40$\% of the hard cross section.  In fact, a 10\% cut on the geometric
cross section corresponds to 30-40\% of the hard cross section for all systems
considered.  Even the central 5\% of collisions encompasses 17-23\% of the hard
cross section while a less stringent cut of 20\% garners 52-66\% of all hard
probes produced before nuclear effects are considered.

\begin{table}[htpb]
\begin{center}
\begin{tabular}{|c|c|c|c|c|} \hline
 & & \multicolumn{3}{c|}{$f_{\rm AB}$} \\
$A+B$   &  $T_{\rm AB}(0)$ (mb$^{-1}$) & $f_{\rm geo} = 0.05$ & $f_{\rm geo} = 
0.1$ & $f_{\rm geo} = 0.2$ \\ \hline
16+16   &  0.79  & 0.166 & 0.306 & 0.522 \\
16+27   &  1.16  & 0.178 & 0.326 & 0.550 \\
16+40   &  1.44  & 0.180 & 0.330 & 0.558 \\
16+63   &  1.89  & 0.185 & 0.340 & 0.576 \\
16+110  &  2.41  & 0.174 & 0.325 & 0.565 \\
16+197  &  3.22  & 0.167 & 0.316 & 0.560 \\
16+208  &  3.18  & 0.164 & 0.312 & 0.555 \\
27+27   &  1.75  & 0.191 & 0.348 & 0.580 \\
27+40   &  2.23  & 0.195 & 0.355 & 0.591 \\
27+63   &  3.00  & 0.201 & 0.365 & 0.606 \\
27+110  &  3.94  & 0.191 & 0.354 & 0.599 \\
27+197  &  5.32  & 0.184 & 0.344 & 0.595 \\
27+208  &  5.27  & 0.181 & 0.340 & 0.591 \\
40+40   &  2.92  & 0.202 & 0.365 & 0.604 \\
40+63   &  4.05  & 0.210 & 0.379 & 0.623 \\
40+110  &  5.53  & 0.204 & 0.371 & 0.619 \\
40+197  &  7.64  & 0.198 & 0.366 & 0.618 \\
40+208  &  7.57  & 0.195 & 0.361 & 0.613 \\
63+63   &  5.77  & 0.220 & 0.395 & 0.642 \\
63+110  &  8.19  & 0.218 & 0.393 & 0.643 \\
63+197  &  11.61 & 0.213 & 0.389 & 0.644 \\
63+208  &  11.57 & 0.211 & 0.385 & 0.640 \\
110+110 &  12.43 & 0.222 & 0.398 & 0.648 \\
110+197 &  18.54 & 0.223 & 0.400 & 0.652 \\
110+208 &  18.64 & 0.221 & 0.397 & 0.649 \\
197+197 &  29.32 & 0.229 & 0.410 & 0.663 \\ 
197+208 &  29.82 & 0.229 & 0.409 & 0.662 \\
208+208 &  30.42 & 0.229 & 0.409 & 0.664 \\  \hline
\end{tabular}
\end{center}
\caption[]{ Values of $T_{\rm AB}(0)$ and the fraction of the hard cross
section for $f_{\rm geo} = 0.05$, 0.1 and 0.2 respectively
in several colliding systems. }
\label{tabtab}
\end{table}

\begin{figure}[htb]
\setlength{\epsfxsize=\textwidth}
\setlength{\epsfysize=0.4\textheight}
\centerline{\epsffile{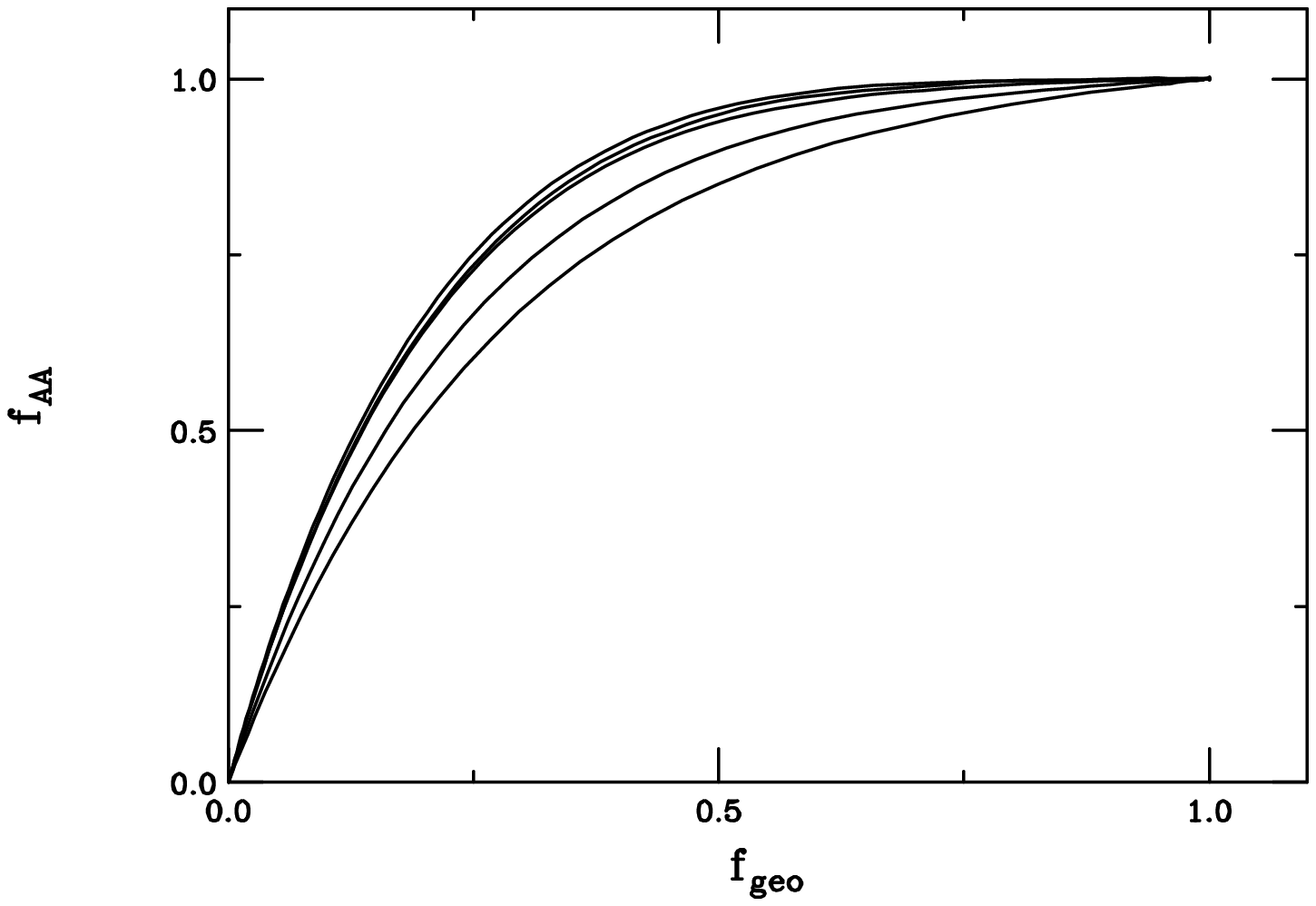}}
\caption[]{The fraction of the hard cross section as a function of the
total geometrical cross section for the symmetric systems shown in Figs.~2 and
3. From left to right, the curves are 197+197, 110+110, 63+63, 27+27, and
16+16.}  
\label{ffsym}
\end{figure}

\begin{figure}[htb]
\setlength{\epsfxsize=\textwidth}
\setlength{\epsfysize=0.4\textheight}
\centerline{\epsffile{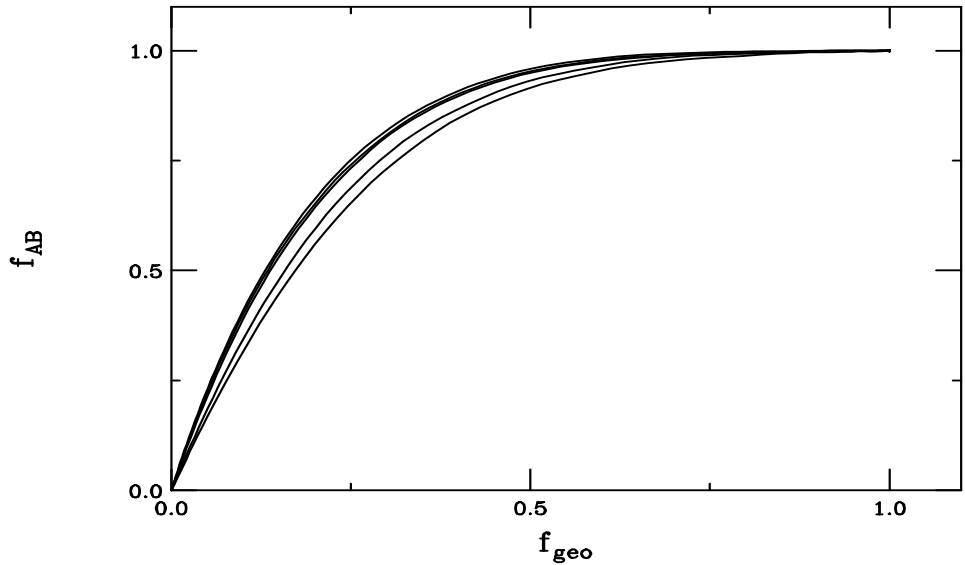}}
\caption[]{The fraction of the hard cross section as a function of the
total geometrical cross section for several asymmetric systems.  From left to 
right, the curves are 197+197, 110+197, 63+197, 27+197, and 16+197.} 
\label{ffasym}
\end{figure}

In Fig.~\ref{ffsym}, 
we show the ratio of $f_{\rm AB}$ relative to the geometric 
ratio for the same systems as in Figs.~\ref{fab} and \ref{fgeo}.  
The hard fraction grows more
slowly relative to the geometric fraction in smaller systems, 16+16 and 27+27,
but otherwise the results are similar.
Figure~\ref{ffasym} shows the 
relative ratios for the asymmetric systems 16+197, 27+197, 63+197, 110+197 
and 197+197.  In this case, the relative ratios cluster even closer together
than for symmetric systems.  Thus the larger nucleus sets the scale for both
the hard and geometric cross sections in asymmetric systems.

\begin{figure}[htb]
\setlength{\epsfxsize=\textwidth}
\setlength{\epsfysize=0.4\textheight}
\centerline{\epsffile{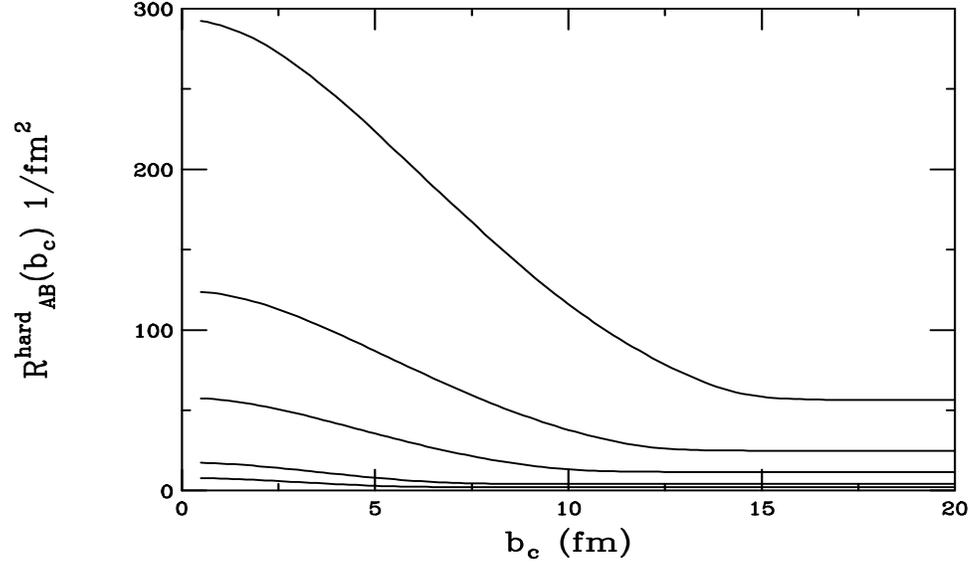}}
\caption[]{The average number of hard probes produced within impact parameter
$b_c$ for the symmetric systems shown in Figs.~2 and
3. From top to bottom, the curves are 197+197, 110+110, 63+63, 27+27, and
16+16.}  
\label{rat}
\end{figure}

The most appropriate way to obtain the number of hard probes
produced in a central collision is to calculate $b_c$ from the geometric cross
section and then, with this $b_c$, calculate $f_{\rm AB}$ with
Eq.~(\ref{abfrac}).  Assuming no nuclear effects on the hard probe production 
cross section, the average hard process rate at $b=0$ in Au+Au collisions is
\begin{eqnarray}
\overline N_{\rm AB}^{\rm hard}(0) = 
\sigma^{\rm hard}_{pp} T_{\rm AB}(0) \, \,
, \label{hardb0} \end{eqnarray}
where $\sigma^{\rm hard}_{pp}$ is the total hard process production
cross section in $pp$ interactions.  The rate in the impact parameter interval
$0<b<b_c$ is the ratio of the hard to geometric cross sections integrated over
$b$,
\begin{eqnarray}
\overline N_{\rm AB}^{\rm hard}(b_c) = \frac{\int_0^{b_c} d\sigma_{\rm 
AB}^{\rm hard}}{\sigma_{\rm geo}(b_c)}
= \frac{\sigma^{\rm hard}_{pp}}{\sigma_{\rm geo}} \frac{AB f_{\rm 
AB}}{f_{\rm geo}} \, \,
, \label{hardbc} \end{eqnarray}
using Eqs.~(\ref{abhard}), (\ref{abfrac}), (\ref{etgeo}), and (\ref{geofrac}).
In Fig.~\ref{rat}, 
the ratio \begin{eqnarray} R_{\rm AB}^{\rm hard}(b_c) \equiv 
\frac{\overline N_{\rm AB}^{\rm hard}(b_c)}{\sigma^{\rm 
hard}_{pp}} = \frac{1}{\sigma_{\rm geo}} 
\frac{AB f_{\rm AB}}{f_{\rm geo}} 
\end{eqnarray} is
shown for the same set of symmetric systems as in Figs.~2--4 as a
function of $b_c$.

As a specific example, the average
number of $c \overline c$ pairs produced at $b=0$ in Au+Au collisions is 
\begin{eqnarray}
\overline N_{\rm AB}^{c \overline c}(0) = 
\sigma^{c \overline c}_{pp} T_{\rm AB}(0) \, \,
. \label{ccb0} \end{eqnarray}
At $\sqrt{s} = 200$ GeV, with MRS D$-^\prime$ parton distributions, 
$\sigma^{c \overline c}_{pp} = 0.344$ mb
\cite{OC} and  $T_{\rm AB}(0) = 29.3$/mb,
resulting in $\approx 10 \,\, c \overline c$ pairs per Au+Au collision at
$b=0$.  With a 10\% centrality cut, the average number of $c \overline c$ pairs
produced in the range $0<b<b_c$ is \begin{eqnarray}
\overline N_{\rm AB}^{c \overline c}(b_c) = \frac{\sigma^{c \overline 
c}_{pp}}{\sigma_{\rm geo}} \frac{AB f_{\rm AB}}{0.1} \, \,
. \label{ccbc} \end{eqnarray} Since the central 10\% of the geometric cross 
section corresponds to 40\% of the hard cross section, there are 
$\approx 7.7\,\, c
\overline c$ pairs in the 10\% most central events.\\
  
{\bf Acknowledgements:}  
I would like to thank A. Morsch, J. Schukraft and P. Jacobs
for helpful discussions. This work was supported in
part by the Division of Nuclear Physics of the Office of High Energy
and Nuclear Physics of the U. S. Department of Energy under Contract
Number DE-AC03-76SF0098.

                        \end{document}